\documentclass[%
 amsmath,amssymb,
raggedfooter,
]{revtex4-2}

\usepackage[
top    = 2.0cm,
bottom = 2.00cm,
left   = 2.00cm,
right  = 2.00cm]{geometry}
    
\usepackage{graphicx}% Include figure files
\usepackage{subfig}
\usepackage{bm}% bold math
\usepackage{xcolor}% bold math
\begin{document}

\title{Sensitivity to decays of long-lived dark photons at the ILC}

\author{Laura Jeanty}
\affiliation{University of Oregon}
\email{Laura.Jeanty@cern.ch}

\author{Laura Nosler}
\affiliation{University of Oregon}

\author{Chris Potter}
\affiliation{University of Oregon}

\date{\today}% It is always \today, today,

\begin{abstract}
We investigate the sensitivity to long-lived dark photons produced in Higgstrahlung events via the Higgs portal, $H \rightarrow \gamma_{D}\gamma_{D}$, with the Silicon Detector (SiD) at the International Linear Collider (ILC). The dark photon model provides a useful benchmark for the ILC sensitivity to light, weakly-coupled new particles that could be identified via a displaced-vertex signature. The ILC is one of several Higgs factories proposed by the international community to study the properties of the Higgs boson at high precision. 
\end{abstract}


\maketitle

\section{Introduction} \label{sec:intro}

Searches for light, weakly coupled particles are an important component of the physics program at present and future colliders. New hidden or dark sectors around the electroweak scale which are weakly coupled to the Standard Model (SM) through mediators are well motivated by numerous theoretical and observational considerations, including naturalness, dark matter, and electroweak baryogenesis. A classic benchmark for a potential vector-boson mediator between the SM and dark sector is the hypothetical dark photon, $\gamma_D$, which interacts with the SM through kinematic mixing with the weak hypercharge field $B$ with coupling strength $\epsilon$. The dark sector could also have a dark Higgs boson, $h_D$, which in the general case will mix with the SM Higgs boson~\cite{Curtin:2014cca}. This opens up a Higgs portal production mode for dark photons.

Prospects for sensitivity to $\gamma_D$ and $h_D$ production have mainly focused on prompt, leptonic decays of the $\gamma_D$ for vector portal $\gamma_D$ production when the mass of the dark photon is greater than about 1~GeV. For small enough $\epsilon$ ($\epsilon \lessapprox 10^{-5}$), the $\gamma_D$ becomes long-lived, a mode which is accessible if Higgs portal production is also considered~\cite{Curtin:2014cca, Ellis:2691414}. The prospects for detection of long-lived particles produced via the Higgs portal at future linear colliders has been studied for displaced hadronic decays, focusing on CEPC and FCC-ee~\cite{Alipour_fard_2019}.

This work addresses the sensitivity for detection of long-lived dark photons at the Silicon Detector (SiD) at the International Linear Collider (ILC). Existing work on $\gamma_D$ and $h_D$ production at the ILC has focused on prompt di-muon decays of $\gamma_D$ production and indirect constraints from measurements of the Higgs to invisible branching ratio~\cite{Ellis:2691414, de_Blas_2020}. We add the Higgs portal production mode and use the displaced decays of long-lived $\gamma_D$ as a benchmark to study the detector performance for detection of displaced decays. We focus on the intermediate $\gamma_D$ masses in the range 0.1-10~GeV. This covers a region which is difficult to access at the colliders due to significant background and which is above the mass reach of fixed target experiments.

In this work, we consider the Higgstrahlung production of $\gamma_D$ via an intermediate $h_{D}$, $e^+ e^- \rightarrow ZH$, followed by $H \rightarrow \gamma_{D} \gamma_{D}$. In Section~\ref{sec:acc}, we study the acceptance to potential displaced $\gamma_{D}$ decays at event generator level. Assuming the selection is tight enough to provide a background free search, we estimate the ILC sensitivity to $\gamma_{D}$ production in terms of the minimum Higgs branching ratio to $h_{D}$ to which SiD would be sensitive in Section~\ref{sec:sensitivity}. In Section~\ref{sec:full_sim}, we outline event selection, reconstruction, and background considerations from full simulation.

\section{ILC and SiD}

\subsection{International Linear Collider: A Higgs Factory}

Since the discovery of the Higgs boson in 2012 \cite{Aad:2012tfa,Chatrchyan:2012ufa} at the Large Hadron Collider (LHC), a consensus has emerged that measuring the properties of the Higgs boson with high precision in \emph{so-called} Higgs factories is an important further step in experimental particle physics. One such Higgs factory is the ILC, one of several proposals currently under consideration by the international community for the next major step in high energy particle physics \cite{fujii2021ilc,Adachi:2022zmw}. 

The ILC is a linear $e^+ e^-$ collider which can operate at $\sqrt{s}=250$~GeV and higher. This nominal $\sqrt{s}$ is close to the maximum for the Higgstrahlung process, $e^+ e^- \rightarrow ZH$, which produces order $10^6$ Higgs bosons at nominal integrated luminosities. Depending on the mixing of the SM Higgs boson $H$ with a dark Higgs boson $h_{D}$, the decay of the Higgs boson can produce dark photon pairs $H \rightarrow \gamma_{D} \gamma_{D}$, which then decay via a kinematic mixing with the SM photon to fermion pairs, $\gamma_D \rightarrow f\bar{f}$. For the $\gamma_D$ masses considered in this study, the couplings to $f\bar{f}$ can be considered equal to the SM photon couplings. If $\epsilon \lessapprox 10^{-5}$, the decays are displaced and can be identified via a signature of a displaced vertex or displaced jet in the vertex and/or calorimeter subdetectors of SiD.

%Relevant to the precision expected for Higgs boson properties at the ILC is the \emph{recoil mass} technique. In contrast to proton colliders, where the colliding protons are not fundamental, in electron-positron colliders the initial state energy can be identified with $\sqrt{s}$ of the beams. Thus, since the ILC  $\sqrt{s}$ can be measured to very high precision, the Higgstrahlung final state $ZH$ energy is also known to high precision. If the $Z$ boson energy $E_{Z}$ is reconstructed then the $H$ boson recoil mass $m_{rec}^2=s-2\sqrt{s}E_{Z}-m_{Z}^2$ cleanly identifies and distinguishes Higgstrahlung events from background events, irrespective of Higgs boson decay.

\subsection{SiD, the Silicon Detector}

Two detector concepts have been proposed for the ILC: SiD and the International Large Detector (ILD). The SiD Detailed Baseline Design (DBD) exercise is documented in the ILC Technical Design Report (TDR) \cite{Behnke:2013lya}. Since publication of the TDR, new detector technologies promise more precise performance for the SiD design, in particular Monolithic Active Pixel Sensors (MAPS) \cite{ballin2009digital}.

SiD is a compact collider detector concept designed for precision all-Silicon vertexing, tracking, and electromagnetic calorimetry. Silicon pixels (strips) make up the sensitive elements in the vertex detector (tracker), both described more fully below. The electromagnetic calorimeter (ECal) is a sampling calorimeter which alternates Tungsten passive layers with Silicon strip active layers, while the hadronic calorimeter (HCal) and muon detector alternate passive Steel layers with plastic scintillator active layers. 

\begin{figure}[t]
\begin{center}
\includegraphics[width=0.65\textwidth]{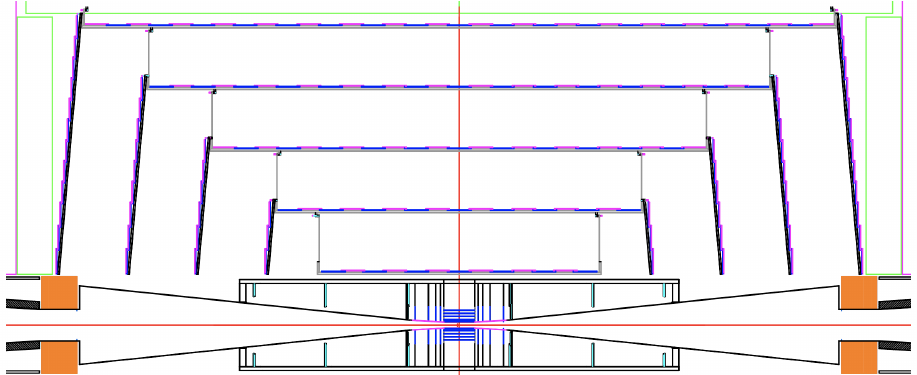}
\caption{The baseline SiD vertex detector and tracker, with the forward beam cones. The vertex detector comprises five barrel layers around the beampipe capped by eight disks. The tracker comprises another five barrel layers capped by eight conical disks. Six forward disks surround the forward cones. The vertex detector (tracker) barrel layers and disks are instrumented with Silicon pixels (strips). Taken from \cite{Behnke:2013lya}.}
\label{fig:sidvtx}
\end{center}
\end{figure}

In a collider detector, reconstruction of displaced decays of long-lived particles to lepton and quark pairs via a displaced-vertex signature places stringent requirements on tracks with large impact parameter and vertexing. Displaced dark photon decays $\gamma_{D} \rightarrow e^+ e^-,\mu^+ \mu^-,\tau^+ \tau^-,q\bar{q}$ require precision spatial distinction between the primary vertex and secondary vertices resulting from long-lived particle decays. See Figure \ref{fig:sidvtx} for the baseline SiD vertex detector and tracker design.

\subsection{Vertex Detector, Tracker and Particle Flow}

The SiD vertex detector comprises five barrel layers surrounding the beampipe at radii from 1.4~cm to 6.0~cm and four disks capping the barrel on each side of the collision point to 17.2~cm along the beamline. Three forward disks surround the forward cones on each side of the collision point to 83.2~cm. The barrel layers and disks in the vertex detector are instrumented with 17~$\mu$m Silicon pixels, yielding 5~$\mu$m hit resolution with less than $0.3\% X_0$ per layer to enable precision vertex determination. The SiD tracker surrounds the vertex detector with another five barrel layers out to a radius of 122.2~cm, each capped by mildly conical disks to 152.2~cm along the beamline on each side of the collision point. The tracker barrel layers and disks are instrumented with Silicon strips. The material budget does not exceed $0.9\% X_0$ ($1.3\% X_0$) in the tracker barrel (disk) layers. For $\theta>20^{\circ}$ the transverse impact parameter uncertainty is expected to be better than $2 \mu$m ($10 \mu$m) for 100 (10) GeV charged particles. 

Critical to the design of SiD is the concept of particle flow, which enables separation and identification of electrons, muons and charged hadrons. By placing the detector solenoid outside of the calorimetry, the tracks reconstructed in the Tracker can be extrapolated through the magnetic field into the calorimetry to match to energy clusters in the ECal and HCal. This allows distinction between electrons and photons in the ECal, and between charged hadrons and neutral hadrons in the HCal. Tracks which do not match to energy clusters in the ECal or HCal can be matched to hits in the muon detector for muon identification. The SiD solenoid is placed between the HCal and the muon detector with a field strength of 5T.

\section{Monte Carlo Simulation Samples}

The production of signal and background simulation samples for this study has been documented in \cite{Potter:2021dkr}. We assume the ILC with $\sqrt{s}=250$~GeV with integrated luminosities of 900 fb$^{-1}$ for each of two polarization cases $e_{L}^{-} e_{R}^{+}$ and $e_{R}^{-}e_{L}^{+}$ at nominal ILC TDR polarization fractions, 80\% electron polarization and 30\% positron polarization. We briefly summarize the production and provide some additional details below.

We use MG5\_aMC@NLO v2.6.6 for signal event generation with events stored in the \texttt{hepmc} format. Events are then processed further with both fast and full simulation. For fast SiD simulation, all signal and background generator samples are passed through the Delphes 3.4.2 \cite{Selvaggi:2014mya,Mertens:2015kba} fast detector simulation with the DSiD detector card \cite{Potter:2016pgp}, though due to the difficulty of simulating displaced decays accurately with fast simulation, only the generator level particle information is used in this study.

\subsection{Model, Benchmarks, and Generator}

We use HAHM dark photon model \cite{hahm} imported to MG5\_aMC@NLO v2.6.6 \cite{2011} for signal generator samples. The four free parameters in this model are the dark photon mass $m_{\gamma_{D}}$, the dark photon mixing parameter $\epsilon$, the dark Higgs mass $m_{h_{D}}$, and the dark Higgs mixing parameter $\kappa$. Because the dark photon lifetime is not determined by the HAHM model, these are calculated first with MG5\_aMC@NLO and inserted manually into the model. For the study at event generation level, samples were generated at masses of 0.1~GeV and from 1 to 10~GeV at 1~GeV increments for both $\epsilon = 10^{-6}$ and $\epsilon = 10^{-7}$. For the full simulation study, see Table \ref{tab:sig} for a summary of the six benchmark signal points used. The decay distance for several $\gamma_D$ masses with $\epsilon = 10^{-6}$ is shown in Figure~\ref{fig:distances}.

\begin{table}[t]
\begin{center}
\begin{tabular}{|c|c|c|c|c|c|} \hline
$m_{\gamma_{D}}$[GeV] & $\epsilon$ & $\Gamma_{\gamma_D}$ [GeV] & $c\tau_{\gamma_D}$ [m] &   $(\beta \gamma) c\tau_{\gamma_D}$ [m] & BR$_{q\bar{q}/\ell^+ \ell^-}$ [\%] \\ \hline \hline
2 & $10^{-5}$ & $2.1 \times 10^{-12}$ & $0.94 \times 10^{-4}$ &  $3.2 \times 10^{-3}$ & 50/50 \\
2 & $10^{-6}$ & $2.1 \times 10^{-14}$  & $0.94 \times 10^{-2}$ &  $3.2 \times 10^{-1}$ & 50/50 \\
2 & $10^{-7}$ & $2.1 \times 10^{-16}$ & $0.94 \times 10^{0}$ &  $3.2 \times 10^{1}$ & 50/50 \\ \hline
10 & $10^{-5}$ & $1.7 \times 10^{-11}$ & $1.2 \times 10^{-5}$ &  $8.5 \times 10^{-5}$ & 62/38 \\
10 & $10^{-6}$ & $1.7 \times 10^{-13}$ & $1.2 \times  10^{-3}$ &  $8.5 \times 10^{-3}$ & 62/38\\ 
10 & $10^{-7}$ & $1.7 \times 10^{-15}$ & $1.2 \times 10^{-1}$ &  $8.5 \times 10^{-1}$ & 62/38 \\ \hline
\end{tabular}
\caption{Dark photon masses $m_{\gamma_D}$ and mixing parameters $\epsilon$ used for signal Monte Carlo event generation with MG5\_aMC@NLO using the HAHM model. Also shown are the dark photon widths $\Gamma_{\gamma_D}$, $c\tau_{\gamma_D}$, mean decay distance $(\beta \gamma) c \tau_{\gamma_D}$, and dark photon branching ratios to hadrons and charged leptons. The boost factors are approximately $\beta \gamma\approx \frac{1}{2}E_{H}/m_{\gamma_D}$. In all cases $m_{h_D}=65$~GeV and $\kappa=7.5 \times 10^{-4}$.}
\label{tab:sig}
\end{center}
\end{table}

We calculate the partial widths for Higgs boson decays to dark photon pairs $H \rightarrow \gamma_D \gamma_D$ and dark photon decays to quark and charged lepton pairs $\gamma_D \rightarrow q\bar{q},\ell^+ \ell^-$ with MG5\_aMC@NLO using the HAHM model. From these partial widths the branching ratios for these decays are calculated. The Higgs boson branching ratios are recalculated as 

\begin{equation}
BR(H \rightarrow \gamma_D \gamma_D)   =  \frac{\Gamma_{H \rightarrow \gamma_D \gamma_D}}{\Gamma_{H}^{SM}+\Gamma_{H \rightarrow \gamma_D \gamma_D}} \\
\end{equation}

\noindent where we assume $\Gamma_{H}^{SM}=4.025$~MeV \cite{LHCHiggsCrossSectionWorkingGroup:2011wcg}. See Tables \ref{tab:sig} and \ref{tab:brdp}.

For signal events, the Higgstrahlung process $e^+ e^- \rightarrow ZH$ at $\sqrt{s}=250$~GeV is specified and the $H$  is required to decay to a pair of dark photons $H \rightarrow \gamma_{D} \gamma_{D}$. The dark photons are then allowed to decay inclusively to fermion pairs $\gamma_{D} \rightarrow f\bar{f}$ according to the branching ratios determined by the HAHM model. For the study at event generation level, separate samples of $\gamma_{D}\rightarrow q\bar{q}$ and $\gamma_{D}\rightarrow l\bar{l}$ are produced. The $Z$ decays hadronically according to the SM branching ratios. Pythia8 \cite{Sjostrand:2007gs} performs the hadronization of quarks in the case of decays to quark pairs. Because MG5\_aMC@NLO does not include initial state radiation (ISR) or beamstrahlung, their impact on this study is assessed in the treatment of systematic uncertainties. 

For each benchmark point in Table \ref{tab:sig}, $5\times 10^4$ events with fully inclusive $\gamma_D$ decays are generated each for the two polarization cases with the nominal 80\% electron 30\% positron polarization fractions, of which $5\times 10^3$ events are fully simulated and reconstructed. Additionally, a second set of identical samples are produced, except with exclusive $\gamma_D$ decay to charged lepton pairs $\gamma_D \rightarrow \ell^+ \ell^-$. Each fully simulated and reconstructed leptonic $\gamma_D$ decay sample contains $5 \times 10^2$ events.

During the DBD exercise generator samples for the ILC with $\sqrt{s}=250$~GeV and pure polarization states $e_{L}^{-} e_{R}^{+}$ and $e_{R}^{-}e_{L}^{+}$ with 100\% polarization fractions were produced in \texttt{stdhep} format with Whizard 1.6 \cite{Kilian:2007gr}. These samples have been preserved and are in use for this study. In particular, samples with Higgstrahlung events with $m_{H}=125$~GeV and inclusive $Z$ and $H$ decays (\texttt{higgs\_ffh}), with polarizations mixed to reflect the nominal ILC beam polarization, are used in this study.

Generic background samples with all nominal SM processes were also produced for the DBD exercise. They include 2-fermion states $e^+e^- \rightarrow f\bar{f}$, 3-fermion states $e\gamma \rightarrow eZ,\nu W \rightarrow 3f$ and 4-fermion states $e^+e^- \rightarrow WW,e\nu W,ZZ, eeZ, \nu \bar{\nu}Z \rightarrow 4f$. In order to obtain a single sample including all SM backgrounds at nominal ILC beam polarization, these samples were mixed by SiD and weighted by cross section, targeting an integrated luminosity of 250 fb$^{-1}$ for each beam polarization case. The resulting sample \texttt{all\_SM\_background} provides a useful indication of likely backgrounds, though some processes have large event weights and therefore present large statistical uncertainty.

\begin{table}[t]
\begin{center}
\begin{tabular}{|c|c|c|c|c|c|c|} \hline 
 & \multicolumn{3}{|c|}{$\Gamma_{H \rightarrow \gamma_D \gamma_D}$ [MeV]} & \multicolumn{3}{|c|}{BR($H \rightarrow \gamma_D \gamma_D$)} \\ 
$m_{h_D}$ [GeV] & $\kappa=0.01$ & $\kappa=0.005$ & $\kappa=0.001$ & $\kappa=0.01$ & $\kappa=0.005$ & $\kappa=0.001$ \\ \hline \hline
10 & 0.470 & 0.118 & 0.00470 & 10.5\% & 2.8\% & 0.1\% \\
20 & 0.489 & 0.122 & 0.00489 & 10.8\% & 2.9\% & 0.1\% \\
50 & 0.658 & 0.165 & 0.00658 & 14.1\% & 3.9\% & 0.2\% \\ \hline
\end{tabular}
\caption{Partial widths and branching ratios for decays of the Higgs boson to dark photon pairs for a variety of dark Higgs masses $m_{h_D}$ and dark Higgs mixing parameters $\kappa$. Partial widths are calculated with MG5\_aMC@NLO using the HAHM model.}
\label{tab:brdp}
\end{center}
\end{table}

\subsection{Full SiD Simulation and Event Reconstruction}

\begin{figure}[t]
\begin{center}
\includegraphics[width=1.\textwidth]{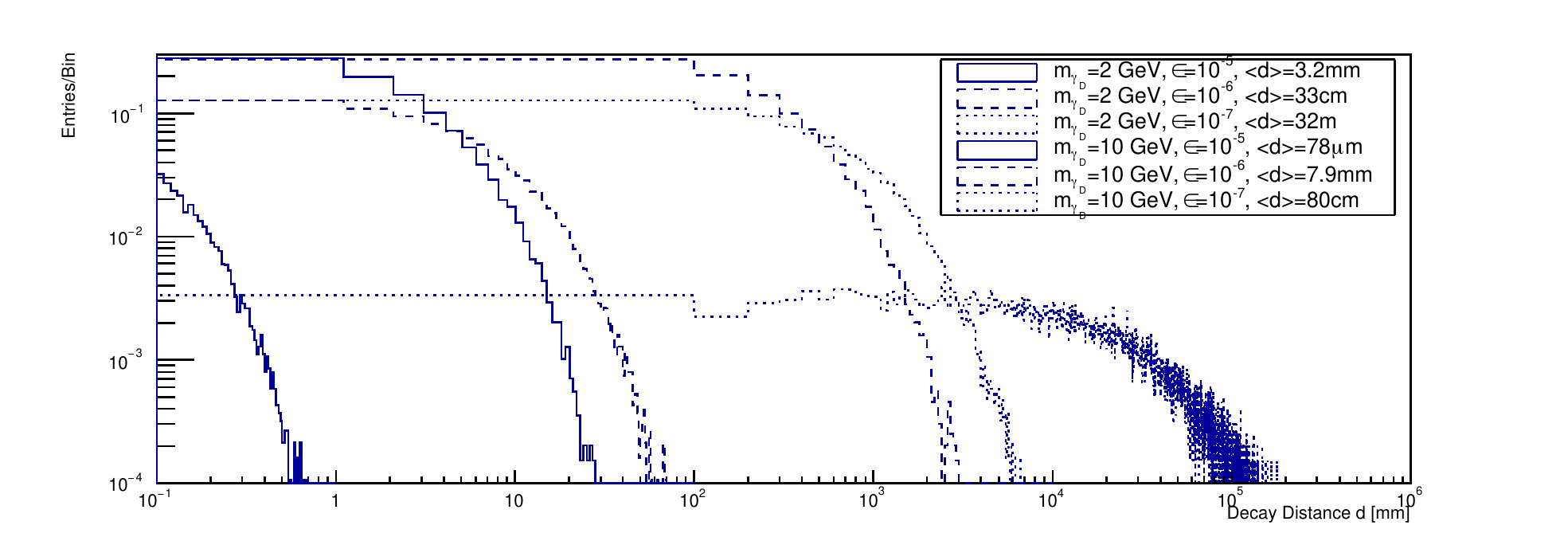}
\caption{The decay distance, normalized to unity, of the $\gamma_{D}$ for the full simulation samples considered in this study.}
%\caption{At left, the decay distance for $\gamma_D$ with $\epsilon = 10^{-6}$ for several different $\gamma_D$ masses. At right, the decay distance, normalized to unity, of the $\gamma_{D}$ for the full simulation samples considered in this study.}
\label{fig:distances}
\end{center}
\end{figure}

For full SiD simulation, all samples are simulated with Geant4 \cite{AGOSTINELLI2003250,Allison:2006ve,ALLISON2016186} using the \texttt{dd4hep} \cite{1742-6596-513-2-022010} interface and the compact SiD detector description option 2, version 3 in the \texttt{lcgeo} package. Background samples were processed in ILCSoft v02-00-02, while signal samples were processed in a local \texttt{dd4hep} and \texttt{lcgeo} build with the \texttt{HepMC3} reader flag enabled to ensure correct treatment of displaced dark photon decay vertices. Because the design of the beampipe and vertex detector support cones impact vertexing performance, an improved SiD beampipe and support cone description has been tested, validated and included on top of the nominal release in this study. See Figure \ref{fig:distances} for the distance of the dark photon $\gamma_{D}$ decay from the interaction point obtained from the generator truth record in the full simulation LCIO samples using the \texttt{getEndpoint} method.

Event reconstruction on full SiD simulation samples is performed with Marlin (Modular Analysis and Reconstruction for the LINear collider) in ILCSoft v02-00-02. Track reconstruction includes track finding, which assigns vertex detector and tracker hits to a candidate track, and track fitting, which determines the helix parameters and from these the charged particle momentum an impact parameters. In this study we employ the Marlin module \texttt{TruthTrackFinder}, which uses MC truth to assign hits to candidate tracks but preserves the track uncertainties from hit measurements in the fitting. While this represents an ideal limit of track finding performance, it effectively factorizes vertex finding performance from track finding performance. 

The nominal package for vertex reconstruction, in use by both ILD and SID, is \texttt{LCFIPlus} \cite{Suehara:2015ura}. \texttt{LCFIPlus} uses a ``tear down'' approach to reconstructing the \emph{primary vertex} at the nominal beamspot, assuming most tracks originate there, and a ``build up'' approach to \emph{secondary vertices}, assuming a small few tracks originate there. To reconstruct the primary vertex candidate, \texttt{LCFIPlus} constrains all tracks in the nominal beamspot, then selectively removes tracks from the candidate primary vertex one-by-one if their vertex $\chi^2$ exceeds a maximum threshold. 

When the primary vertex candidate $\chi^2$ is acceptably low, \texttt{LCFIPlus} then builds candidate secondary vertices using tracks rejected from the primary vertex. All possible track pairs are formed with these rejected tracks and kept as secondary vertex candidates if the pair $\chi^2$ does not exceed a maximum threshold. Additional tracks are then added to the secondary vertex candidates if their $\chi^2$ contribution is small, and a final selection is made to reduce the set of secondary vertex candidates to one with unique track assignment.

\section{Dark Photon Signal Acceptance at SiD}
\label{sec:acc}

Displaced decays of $\gamma_D$ can be selected with the reconstruction and identification of a displaced vertex. If the $\gamma_D$ decays before the end of the vertex detector, tracks with large impact parameter and a displaced vertex can be reconstructed using the last layer of the vertex detector and the Tracker. If the $\gamma_D$ decays before the start of the calorimeter, the calorimeter can be used to reconstruct displaced hadrons or leptons. We define two fiducial regions, $R1$ and $R2$, in which we calculate the acceptance for finding a displaced vertex with either the tracker or the calorimeters. 

\begin{figure}[t]
\centering
\subfloat[The mean number of charged particles per $\gamma_D$ decay] {\includegraphics[width=0.45\textwidth]{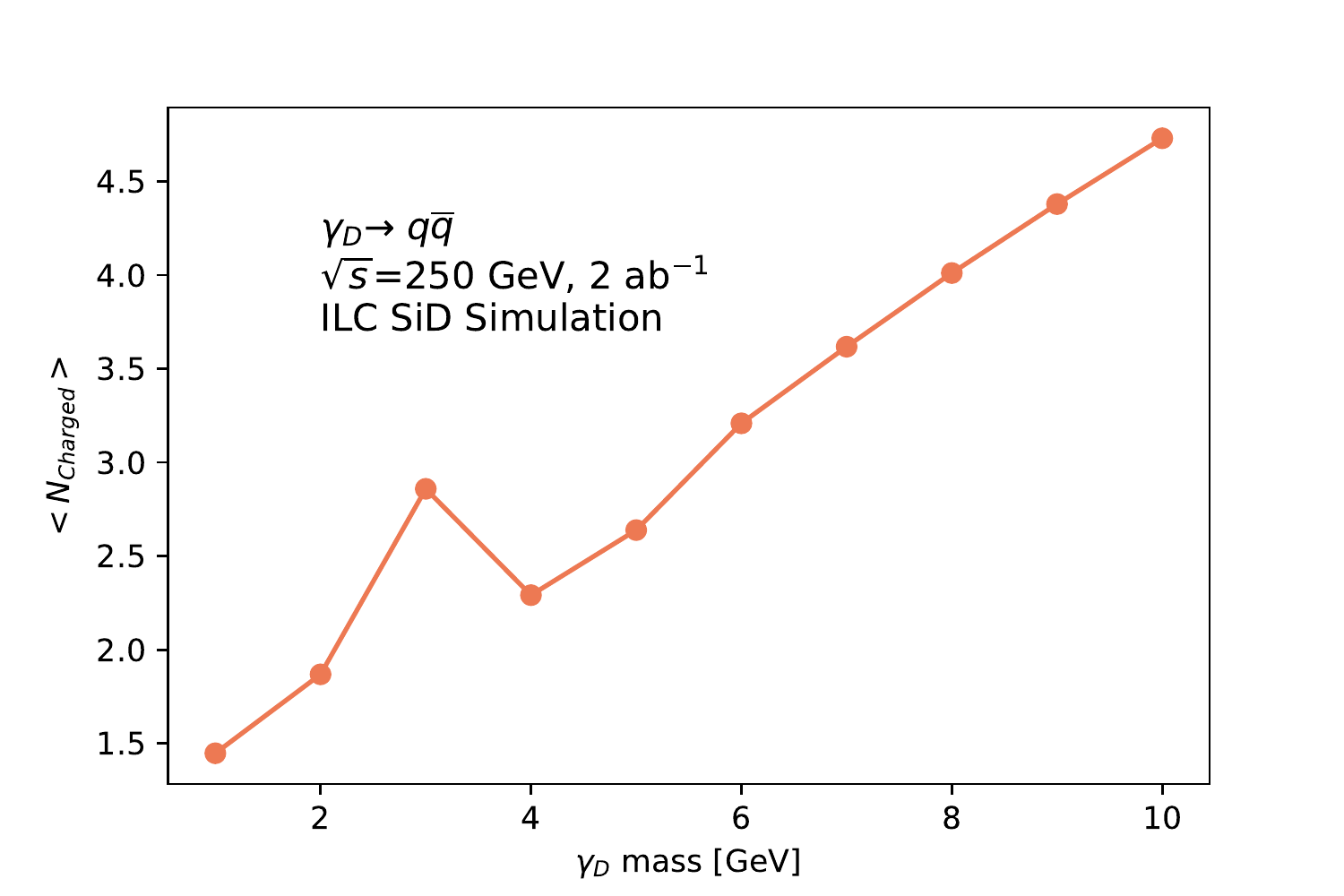}}
\subfloat[The mean momentum of charged particles per $\gamma_D$ decay] {\includegraphics[width=0.45\textwidth]{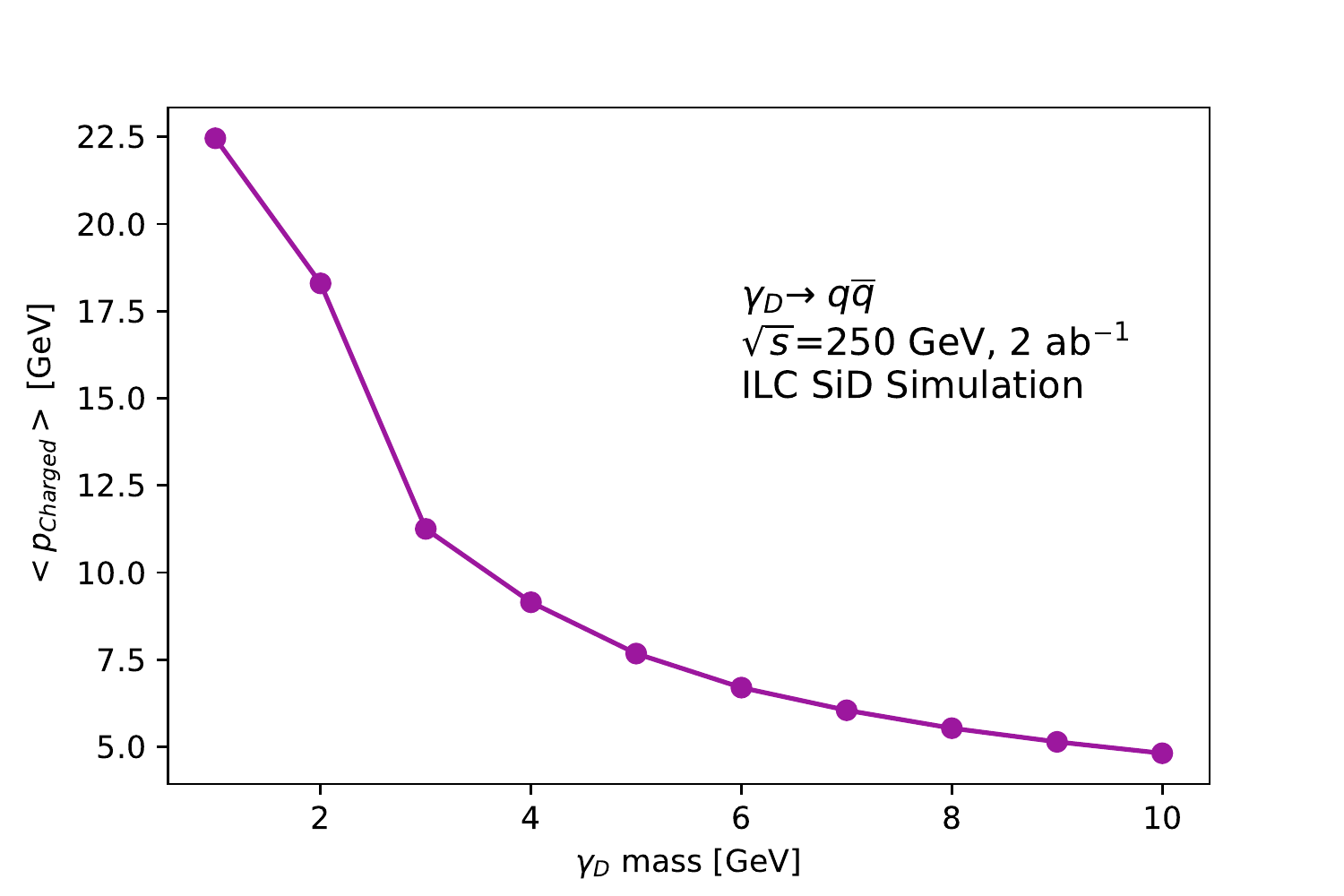}}
\caption{Generator level properties of hadronic $\gamma_D$ decays, as a function of $\gamma_D$ mass. }
\label{fig:avg_hadronic} 
\end{figure}

Acceptances are calculated separately for leptonic and hadronic event selection. To suppress background from hadronic interactions of SM particles, decays of long-lived SM hadrons, and photon conversions, in the case of hadronic selection, we require that the dark photon decay produce at least three charged particles. The mean number of charged particles per $\gamma_D$ decay is shown in Figure~\ref{fig:avg_hadronic}, as is the mean momentum per charged particle, both as a function of $\gamma_D$ mass. The mean number of charged hadrons rises from around 1.5 to above 4.5 as the $\gamma_D$ mass increases from 1 to 10~GeV. Around 3~GeV, the production of $J\psi$s opens up, which causes an increase the number of charged particles from the $\gamma_D$ decay until the mass of the $\gamma_D$ is above the threshold to produce $D$-mesons. The momentum, $p$, per charged particle drops from roughly 22 to 5~GeV as the $\gamma_D$ mass increases from 1 to 10~GeV. For all masses, the decay products of the $\gamma_D$ get a significant boost from the momentum the $\gamma_D$ inherits from the $h_D$, and therefore it should be possible to efficiently select tracks from all dark photons in this mass range, given this production mode.

The requirements for fiducial region $R1$ are:
\begin{itemize}
    \item At least one $\gamma_D$ decays between 2 and 60 mm 
    \item At least 2 leptons, or at least 3 charged hadrons, from the $\gamma_D$ decay must satisfy:
    \begin{itemize}
        \item $p>100$ MeV
        \item $\theta > 20$ degrees
        \item Transverse impact parameter, $d_0$, $> 2$ mm
    \end{itemize}
\end{itemize}

At generator level there is no difference between requiring 3 or 4 charged hadrons from the $\gamma_D$ decay as there are an even number of charged hadrons at each decay. The requirements on $d_0$ and the $\gamma_D$ decay distance are included to suppress background from $B$-hadrons and are purposely conservative in order to reduce easily the background to a negligible level. The requirement on $\theta$ is motivated by the geometry of SiD. While SiD should have efficient pion reconstruction above 100 MeV, as shown in Figure~\ref{fig:avg_hadronic}, the $p$ selection could be tightened significantly with little loss in acceptance. To suppress background from photon conversions, a requirement on the invariant mass of the lepton pair should be imposed as well, but at generator level the acceptance for this selection should be 100\% and therefore it is only interesting to study the resolution of the vertex mass in fully simulated and reconstructed samples.

The requirements for fiducial region $R2$ are similar, with a slightly higher momentum requirement to reflect that the decay products must be reconstructable in the calorimeter, and with a larger decay volume:
\begin{itemize}
    \item At least one $\gamma_D$ decays between 2 and 1250 mm 
    \item At least 2 leptons, or at least 3 charged hadrons, from the $\gamma_D$ decay must satisfy:
    \begin{itemize}
        \item $p > 1$ GeV
        \item $\theta > 20$ degrees
        \item Transverse impact parameter, $d_0$, $> 2$ mm
    \end{itemize}
\end{itemize}

\begin{figure}[t]
\begin{center}
\subfloat[leptonic decays]
{\includegraphics[width=0.45\textwidth]{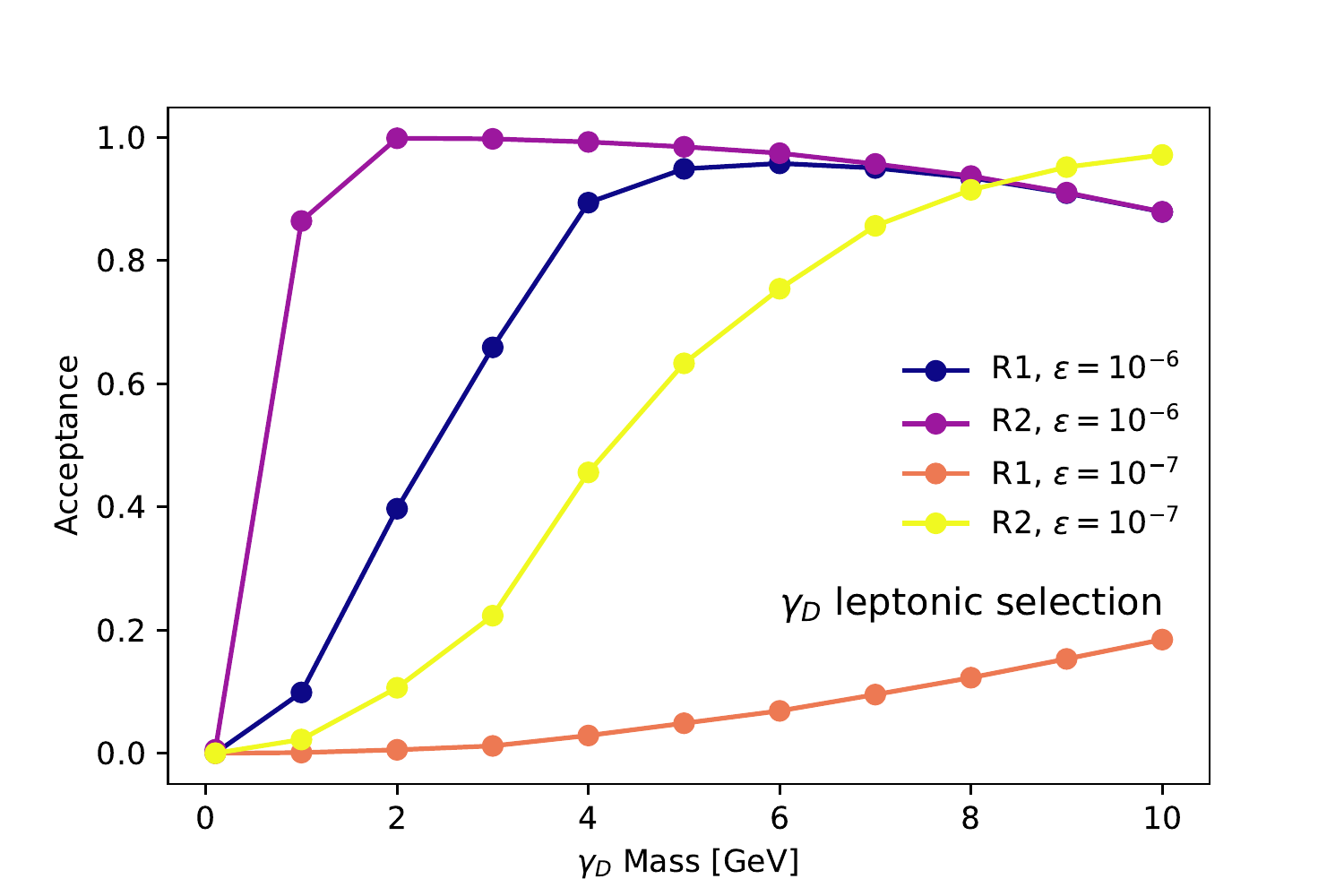}}
\subfloat[hadronic decays] {\includegraphics[width=0.45\textwidth]{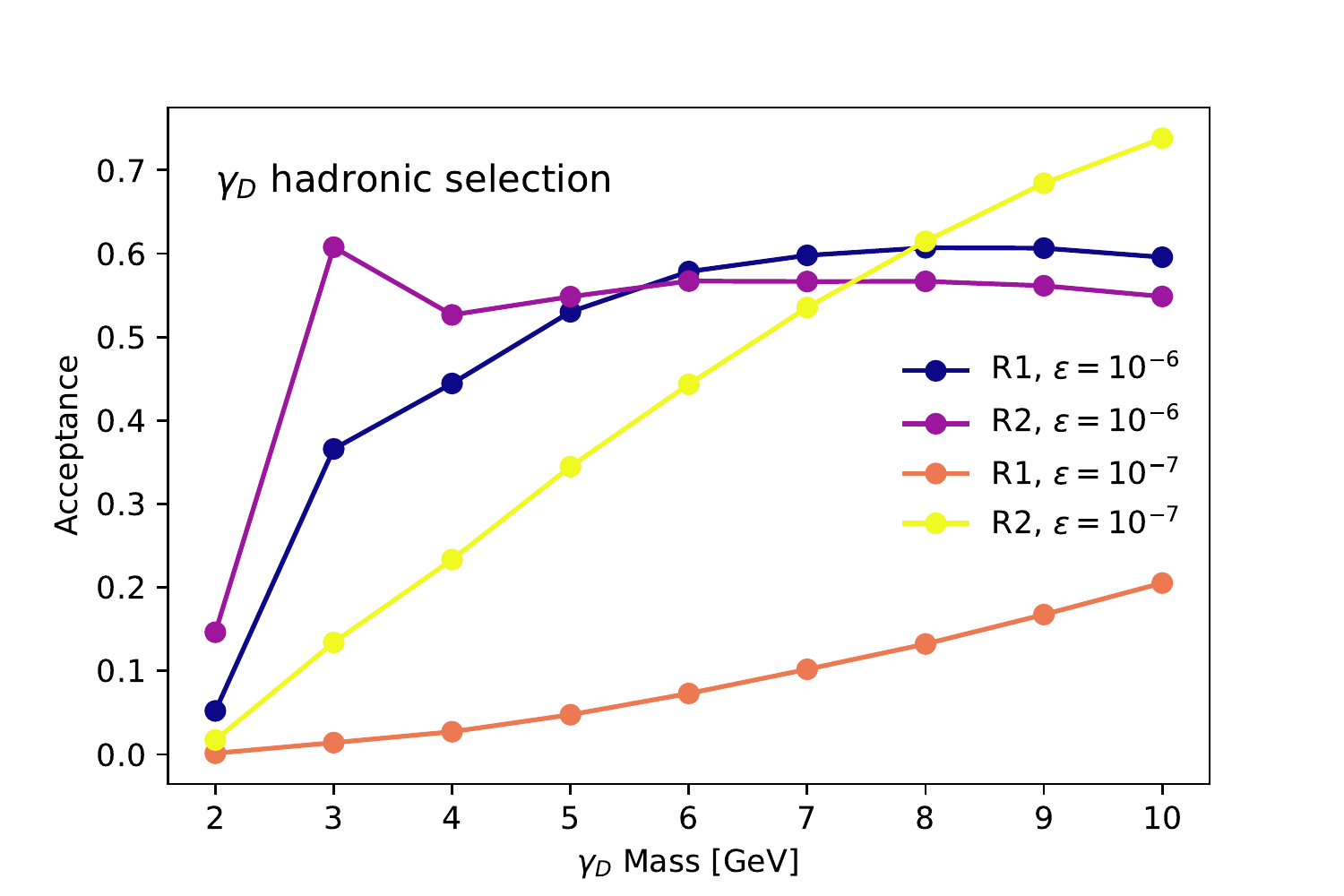}}
\end{center}
\caption{The acceptance for $\gamma_D$ decays in fidicual regions $R1$ and $R2$, for $\epsilon=10^{-6},10^{-7}$, as a function of $\gamma_D$ mass. The lowest mass point in the leptonic selection is shown at 0.1~GeV.}
\label{fig:acceptances}
\end{figure}

\begin{figure}[t]
\centering
\subfloat[leptonic $\gamma_D$ decays]
{\includegraphics[width=0.45\textwidth]{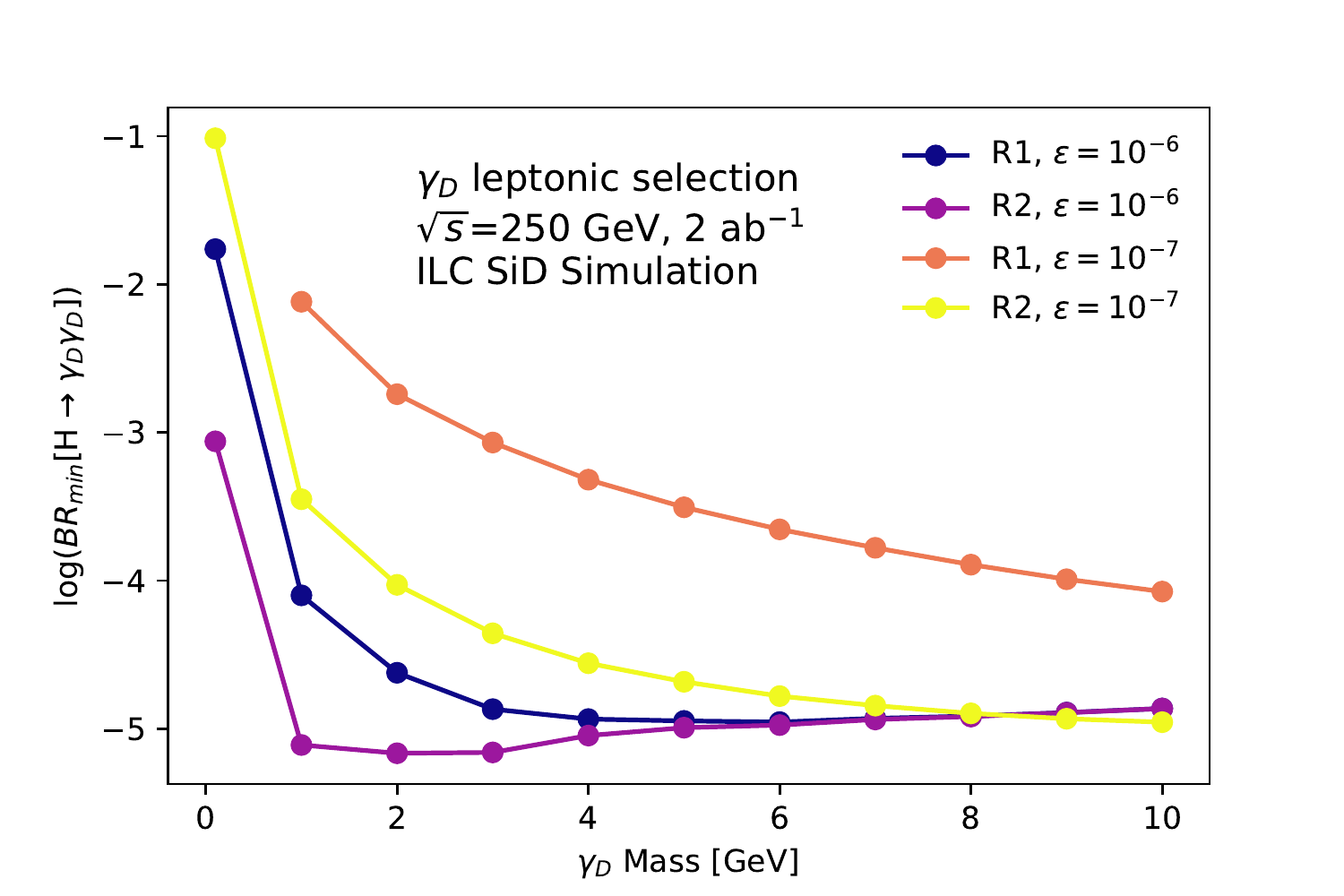}}
\subfloat[hadronic $\gamma_D$ decays] {\includegraphics[width=0.45\textwidth]{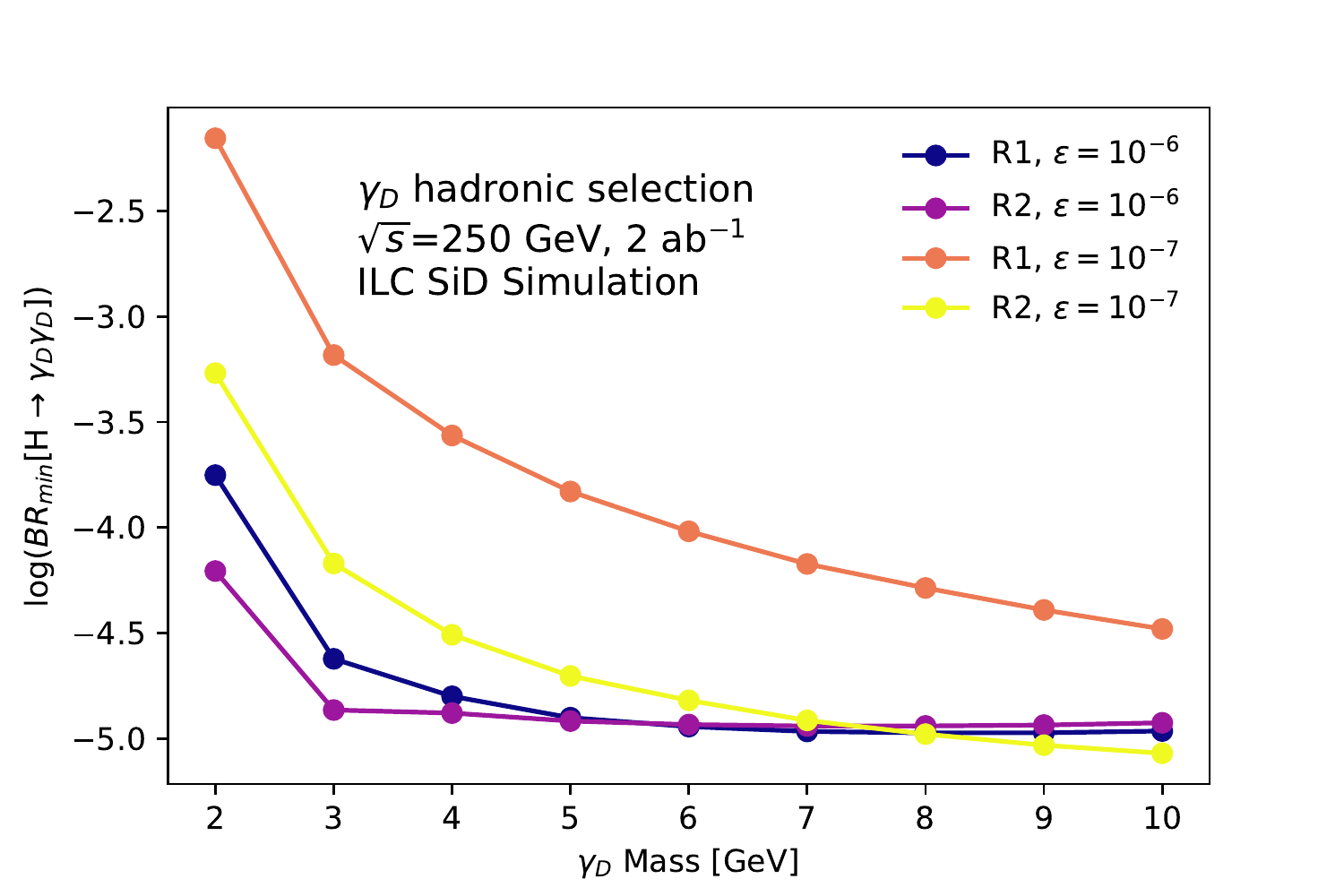}}
\caption{The minimum branching ratio $H \rightarrow \gamma_{D} \gamma_{D}$ to which SiD will be sensitive for $\sqrt{s} = 250$ GeV and 2 ab$^{-1}$.}
\label{fig:minBRs}
\end{figure}

\begin{figure}[t]
\centering
\includegraphics[width=0.65\textwidth]{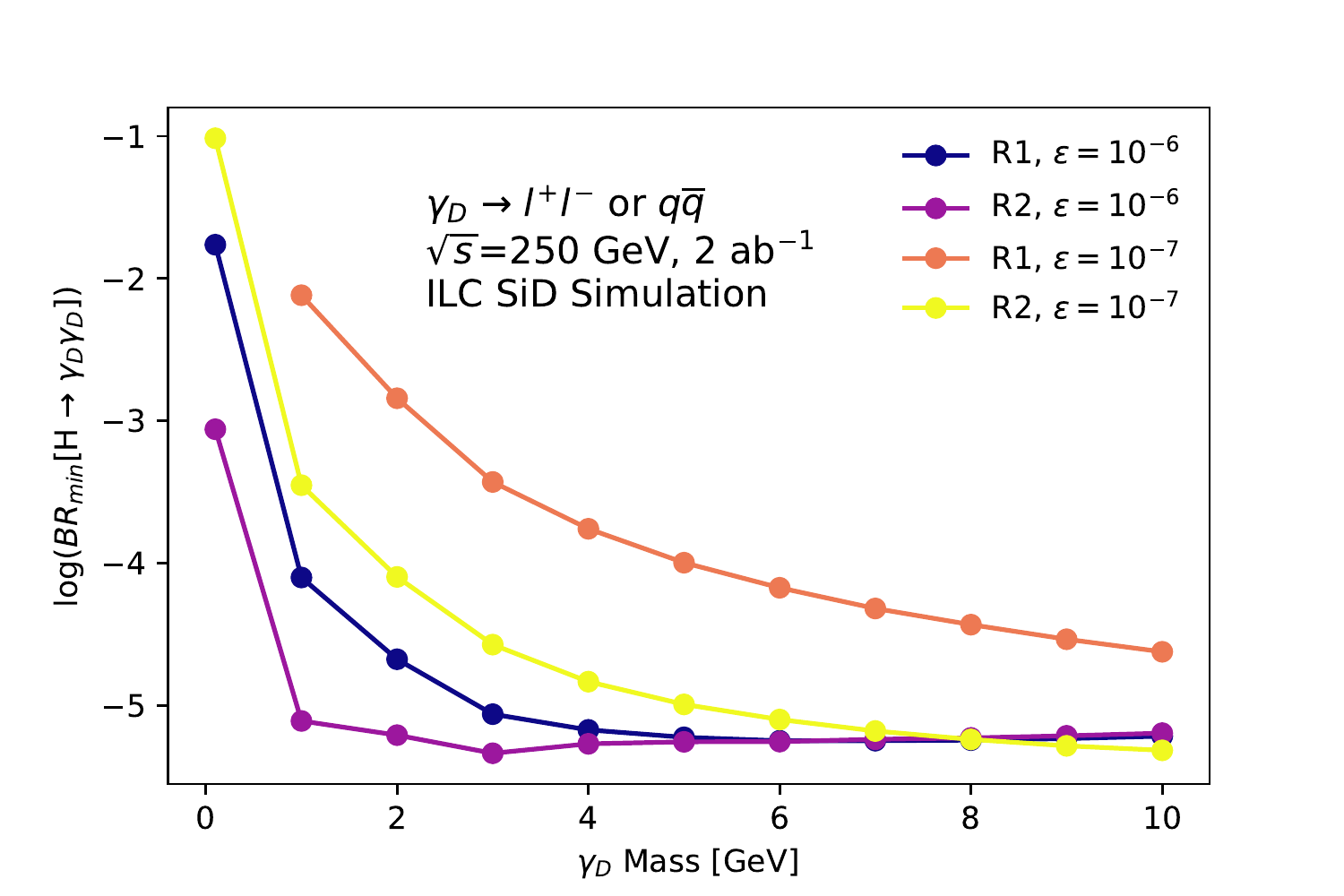}
\caption{The minimum branching ratio $H \rightarrow \gamma_{D} \gamma_{D}$ to which SiD will be sensitive for $\sqrt{s} = 250$ GeV and 2 ab$^{-1}$, when both leptonic and hadronic decays are reconstructed within the regions $R1$ and $R2$, for $\epsilon=10^{-6},10^{-7}$.}
\label{fig:minBR}
\end{figure}

The acceptances for both leptonic and hadronic decays in the two fidicual regions are shown in Figure~\ref{fig:acceptances}. For leptonic decays, the difference in acceptance as a function of $\gamma_D$ mass is driven by the different $c\tau$ that correspond to the same epsilon. For short $c\tau$ corresponding to $\epsilon = 10^{-6}$, the acceptance in $R1$ rises from about 10\% at $\gamma_D$ mass of 1 GeV to nearly 20\% at 10 GeV. With the large fiducial decay volume of $R2$, the acceptance is significantly higher and ranges from 80\% to nearly 100\%. For leptonic decays and $\epsilon = 10^{-7}$, the acceptance smoothly increases from a few percent at 1~GeV to nearly 100\% at 10~GeV. For hadronic decays, the effect on acceptance from the difference in decay position corresponding to different values of $c\tau$ is convoluted with the impact of the variation in charged particle multiplicity.

\section{Baseline ILC Sensitivity to Long-Lived Dark Photons}
\label{sec:sensitivity}

The requirement of a displaced vertex formed from tracks with measurably large impact parameter in the clean ILC event environment likely suppresses background events to a negligible level. Therefore, for a baseline sensitivity, we assume no background with the selections outlined in the previous selection. We also assume that no veto of vertices consistent with the location of detector material is needed for the hadronic decays.

Using the acceptances calculated in the previous section, the production cross-section of $e^+ e^- \rightarrow ZH$ at $\sqrt{s} = 250$~GeV, the branching ratios of $\gamma_{D}$ to quarks and leptons from Ref~\cite{Curtin_2015}, the assumption of no background and full reconstruction efficiency, and the requirement that at least 3 events remain, the minimum branching ratio $H \rightarrow \gamma_{D} \gamma_{D}$ to which SiD will be sensitive is calculated for the two different regions and for leptonic and hadronic selections separately. The minimum branching ratios $H \rightarrow \gamma_{D} \gamma_{D}$ are shown in Figure~\ref{fig:minBRs}, separated by leptonic or hadronic selection. Assuming both decay modes are accessible, the combined sensitivity to both leptonic and hadronic decay modes in shown in Figure~\ref{fig:minBR}.

If both leptonic and hadronic displaced decays can be efficiently reconstructed using the calorimeters, SiD should be sensitive to $\gamma_{D}$ production via Higgstrahlung down to BR$_{H \rightarrow \gamma_{D} \gamma_{D}} = 10^{-5}$ for masses from 1 to 10~GeV and for $\epsilon = 10^{-6}$. This direct search provides significantly more reach than the indirect $H\rightarrow$ invisible search would provide for this new physics scenario. At 1~GeV, this sensitivity extends beyond the projected reach of the HL-LHC~\cite{Curtin_2015} for this particular model.

\section{Event selection and Background Considerations from Full Simulation }
\label{sec:full_sim}

In order to identify backgrounds to the signal $e^+ e^- \rightarrow ZH \rightarrow f \bar{f} \gamma_D \gamma_D$, we first consider SiD detector acceptance regions R1 and R2 and then describe the signal selection. After prompt decay of the $Z$ boson to fermion pairs, $Z \rightarrow f \bar{f}$, and non-prompt decay of the dark photons, $\gamma_D \rightarrow q\bar{q},\ell^+ \ell^-$, there remains a 6-fermion final state. The reconstructed lepton $\ell=e,\mu$ and Particle Flow Object (PFO) multiplicities define the signal topologies.

Displaced particle decays which do not occur in the SM present challenges to track reconstruction, vertex finding and lepton identification with particle flow which SM particle decays do not. For example, if the decay occurs outside of the Vertex Detector (R1) then finding secondary vertices is extraordinarily challenging. If the decay occurs outside of the Tracker (R2), then no tracks will be reconstructed with which to find secondary vertices, and identification of electrons and muons using particle flow will fail. Before describing the signal selection for the full simulation analysis, we consider its motivation from generic considerations of various detector acceptance regions.

\subsection{Acceptance Regions R1 and R2}

We first define four regions based on subdetector sensitivity to $\gamma_D$ decay product detection: region V (vertex detector), region T (tracker), region C (ECal or HCal), and region X (outside HCal). Then region R1 is approximately partitioned into subregions VV, VT, VC, VX based on the decay point of each of the two $\gamma_D$ from the $H \rightarrow \gamma_D \gamma_D$ decay. Region R2 is approximately partitioned into subregions TT, TC and TX. 

In the R1 region, the VV subregion may yield up to two secondary vertices, reconstructed tracks from all charged particles in both $\gamma_D$ decays, and energy deposits in the calorimetry which define the PFOs. The VT subregion is the same, except that only up to one secondary vertex may be reconstructed. In the VC subregion charged particles from one $\gamma_D$ decay are not reconstructed in the tracker, but do leave energy deposits in the calorimetry which define PFOs. These PFOs may fail electron and muon identification and be misidentified as neutrals with degraded energy resolution, but the electrons (if not the muons) will nevertheless be included in the jets. The VX subregion is not considered here.

In the R2 region, the TT subregion yields no secondary vertices but tracks from all charged particles from both $\gamma_D$ decays and energy deposits in the calorimetry defining the PFOs. In the TC subregion tracks from only one $\gamma_D$ decay are reconstructed, but energy deposits in the calorimetry define the PFOs, even if some PFOs fail electron or muon identification and are misidentified as neutral with degraded energy resolution. The TX subregion is not considered here.

For both benchmark points with $\epsilon=10^{-5}$, virtually all events lie in the VV subregion. This is also the case for the $\epsilon=10^{-6}$, $m_{\gamma_D}=10$~GeV point. For the $\epsilon=10^{-6}$, $m_{\gamma_D}=2$~GeV point, approximately 1/3 of events lie in the VV or VT subregion and 2/3 lie in the TT subregion. For the  $\epsilon=10^{-7}$, $m_{\gamma_D}=10$~GeV point, only approximately 1/10 of events lie in the VV, VT or VC subregions where a secondary vertex might be reconstructed. Virtually all events for the $\epsilon=10^{-7}$, $m_{\gamma_D}=2$~GeV point fall in the XX subregion. 

For a given benchmark point in this study, the track, lepton and PFO multiplicities may vary between the five subregions (VV, VT, VC, TT, TC). For different benchmark points, the track, lepton and PFO multiplicities vary even within one subregion. Therefore we make the signal selection completely inclusive for all subregions and benchmark points by chosing the track and PFO multiplicity minima and maxima to be nearly 100\% efficient for all signal events in this study. This procedure still gives strong background rejection for these requirements. In the VC and TC subregions electrons and muons may fail identification with particle flow due to charged particles unreconstructed as tracks, but these events can still be recovered by selecting for a topology with fewer leptons.

The remaining subregions (VX, TX, CX, XX) are not considered in this study, though VX, TX and CX can, in principle, be selected for with a signal selection based on missing energy due to the $\gamma_D$ decay in region X. The subregion XX defines collider invisible Higgs boson decay, which is considered in \cite{https://doi.org/10.48550/arxiv.2203.08330}. 

\subsection{Signal Selection}

We do not consider the cases of leptonic $Z$ decays, $Z \rightarrow \nu \bar{\nu}, \ell^+ \ell^-$ because zero or two tracks in a primary vertex constrained to the beamspot present technical challenges to vertex finding which need further study. We consider only hadronic $Z$ decays $Z \rightarrow q \bar{q}$, which accounts for 70\% of signal events. 

The signal topologies we consider are defined by jet and lepton multiplicity: $2j4\ell$, $4j2\ell$ and $6j$, where the invariant mass of one pair of jets $m_{jj}$ is consistent with the $Z$ mass. We explicitly allow for additional leptons from $\gamma_D \rightarrow q\bar{q}$ or $Z \rightarrow q\bar{q}$ with semileptonic meson decay. The signal selections for the three signal topology cases are as follows:

\begin{itemize}

\item \textbf{Case $2j4\ell$}. Require four or more leptons $N_{e10}+N_{\mu 10} \geq 4$ with $p_{\ell}>10$~GeV. Force the remaining PFOs to two jets. The dijet invariant mass $m_{jj}$ must be consistent with the $Z$ mass. Require either $N_e=4$ with $Q_{4e}=0$, $N_{4\mu}=4$ with $Q_{4\mu}=0$, or $N_e=N_{\mu}=2$ with $Q_{2e}=0$ and $Q_{2\mu}=0$. In the latter case there is no ambiguity in selecting the $\gamma_{D}$ candidates. In the former cases there is a two-fold or worse ambiguity, which is resolved by pairing the leptons such that the difference in candidate $\gamma_D$ invariant masses $m_{\ell^+ \ell^-}$ is minimized.

\item \textbf{Case $4j2\ell$}. Require two or three leptons $N_{e10}+N_{\mu 10}=2,3$ with $p_{\ell}>10$~GeV and $Q_{2\ell}=0$ to form the $\gamma_D \rightarrow \ell^+ \ell^-$ candidate. Force the remaining PFOs to four jets. There is a six-fold ambiguity in jet pairing, which is resolved by choosing the jet pairs such that $\chi^2=(m_{jj}-m_Z)^2/\sigma_{Z}^{2}+(m_{2j2\ell}-m_H)^2/\sigma_{H}^{2}$ is minimized subject to the independent constraint that the mass in recoil from the $Z$ boson candidate is consistent with the $H$ boson mass.

\item \textbf{Case $6j$}. Require zero or one leptons $N_{e10} + N_{\mu 10}\leq 1$. Force all PFOs to six jets. There is a fifteen-fold ambiguity in jet pairing, which is resolved by choosing the jet pairs such that $\chi^2=(m_{jj}-m_Z)^2/\sigma_{Z}^{2}+(m_{4j}-m_H)^2/\sigma_{H}^{2}$ is minimized subject to the independent constraint that the mass in recoil from the $Z$ boson candidate is consistent with the $H$ boson mass. The six-fold ambiguity in determining the $\gamma_D$ candidates is resolved by choosing the pairs such that the difference in invariant mass $m_{jj}$ is minimized.

\end{itemize}

\noindent We use the Durham jetfinder and vary the $y_{cut}$ parameter, a measure of jet separation, to force the jet multiplicity to the desired signal topology.

In all three cases the track and PFO multiplicities must be consistent with signal events. Moreover the jet pair in the hadronic $Z$ boson candidate must be consistent with the signal $Z$ decay opening angle and the known $Z$ mass. The mass in recoil from the $Z$ candidate ($m_{rec}^2=s-2\sqrt{s}E_{jj}-m_{Z}^2$) must be consistent with the $H$ mass. The $H$ boson candidate must be consistent with the signal $H$ decay opening angle and the known $H$ mass. Finally, any $\gamma_D$ candidate, whether hadronic or leptonic, must survive the $K_S$ and $Z$ mass vetoes. 

\begin{figure}[t]
\begin{center}
\includegraphics[width=1.0\textwidth]{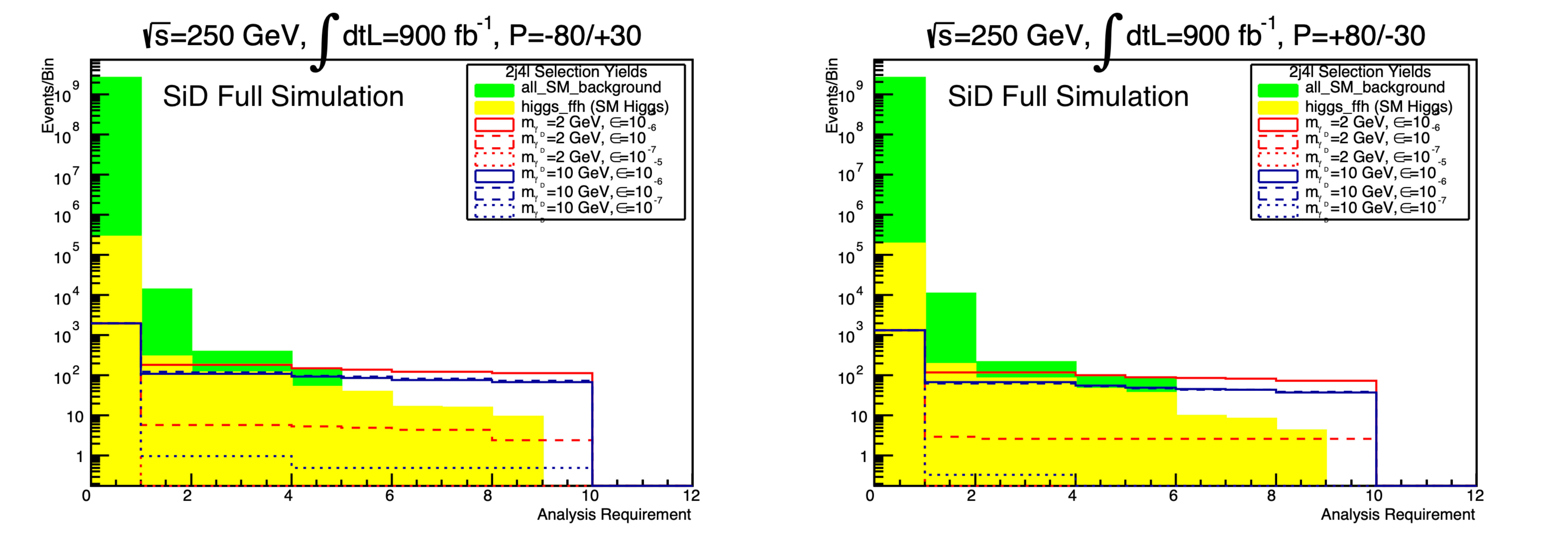}
\end{center}
\caption{Signal and background yields at each step in the $2j4\ell$ signal selection for the $e_{L}^{-} e_{R}^{+}$ (left) and $e_{R}^{-} e_{L}^{+}$ (right) samples at nominal beam polarization fractions. Signal is normalized with $BR(H \rightarrow \gamma_D \gamma_D)=0.01$.}
\label{fig:cutflow}
\end{figure}

Quantitatively, these requirements are as follows. First, the track, lepton, PFO and jet multiplicities must be consistent with signal events:

\begin{itemize}

\item Lepton multiplicity: $N_{\ell} \geq 4, 2 \leq N_{\ell} \leq 3, N_{\ell} \leq 1$ (cases $2j4\ell$,$4j2\ell$,$6j$ respectively)

\item Track/PFO multiplicity: $10 \leq N_{trk} \leq 40$ and $20 \leq N_{pfo} \leq 80$ (all cases)

\item Jet multiplicity: $N_{jet}=2,4,6$ (cases $2j4\ell$,$4j2\ell$,$6j$ respectively)

\end{itemize}

\noindent The track and PFO multiplicity minima and maxima are chosen to be nearly 100\% efficient for all signal events considered in this study (see below). Next, the reconstructed hadronic $Z \rightarrow q \bar{q}$ decay must be consistent with signal Higgstrahlung events:

\begin{itemize}

\item Signal $Z$ candidate decay: $-0.9 \leq \cos \theta_{jj} \leq -0.2$

\item Signal $Z$ candidate mass: $75 \leq m_{jj} \leq 105$~GeV

\item Signal $H$ recoil mass: $110 \leq m_{rec} \leq 140$~GeV

\end{itemize}

\noindent Finally, the reconstructed $H \rightarrow \gamma_D \gamma_D$ decay must be consistent with signal dark Higgs events:

\begin{itemize}

\item Signal $H$ candidate decay: $-0.9 \leq \cos \theta_{2\ell 2\ell,2\ell 2j,2j2j} \leq -0.5$ (cases $2j4\ell$,$4j2\ell$,$6j$ respectively)

\item Signal $H$ candidate mass: $110 \leq m_{4\ell,2\ell 2j,4j} \leq 140$~GeV (cases $2j4\ell$,$4j2\ell$,$6j$ respectively)

\item Signal $\gamma_D$ candidate mass ($K_S,Z$ veto):  $0.6 \leq m_{\ell^+ \ell^-} \leq 10.5$~GeV, $0.6 \leq m_{jj} \leq 12.5$~GeV 

\end{itemize}

\begin{figure}[t]
\begin{center}
\includegraphics[width=1.0\textwidth]{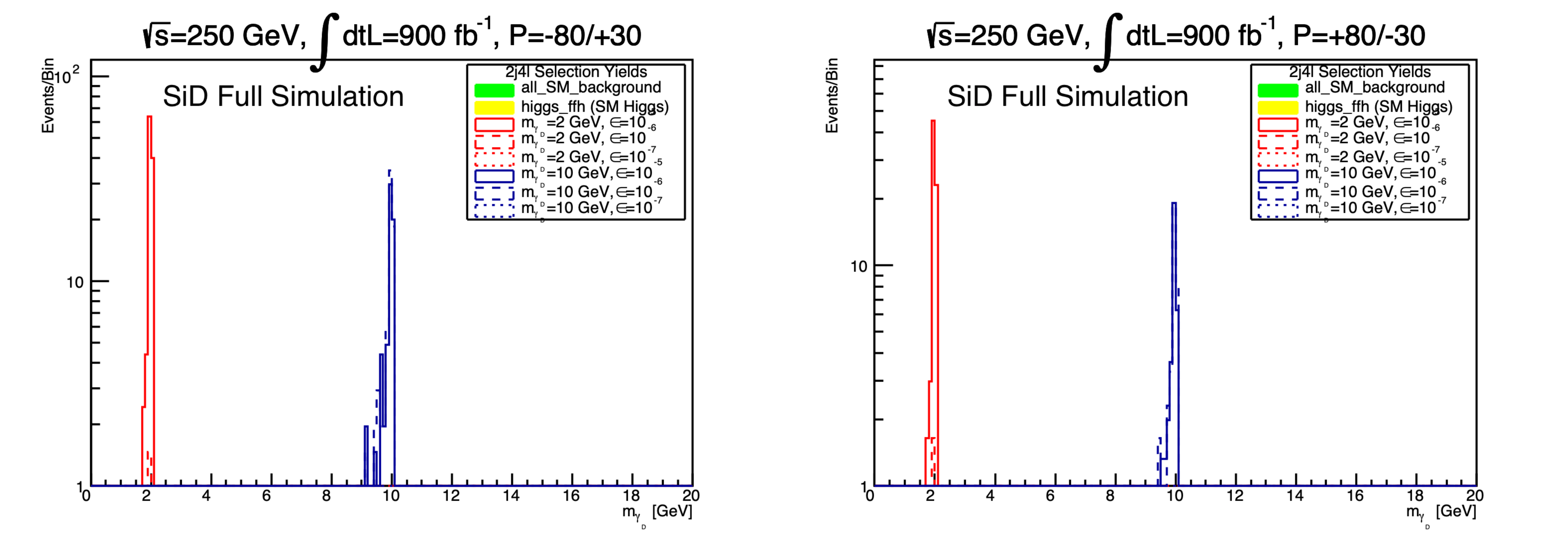}
\end{center}
\caption{Candidate $\gamma_D \rightarrow \ell^+ \ell^-$ mass after full $2j4\ell$ signal selection for the $e_{L}^{-} e_{R}^{+}$ (left) and $e_{R}^{-} e_{L}^{+}$ (right) samples at nominal beam polarization fractions. Signal is normalized with $BR(H \rightarrow \gamma_D \gamma_D)=0.01$. }
\label{fig:zdmass}
\end{figure}

\noindent The hadronic $\gamma_D$ candidate mass maximum is looser than the leptonic one to allow for the lower hadronic mass resolution. See Figure \ref{fig:cutflow} for the yields in the $2j4\ell$ signal selection cut flow and Figure \ref{fig:zdmass} for the candidate dark photon $\gamma_D \rightarrow \ell^+ \ell^-$ mass after full signal selection. For the background yields at each stage in the $2j4\ell$, $4j2\ell$ and $6j$ signal selection, see Table \ref{tab:bgyields}. 

\begin{figure}[t]
\begin{center}
\includegraphics[width=1.0\textwidth]{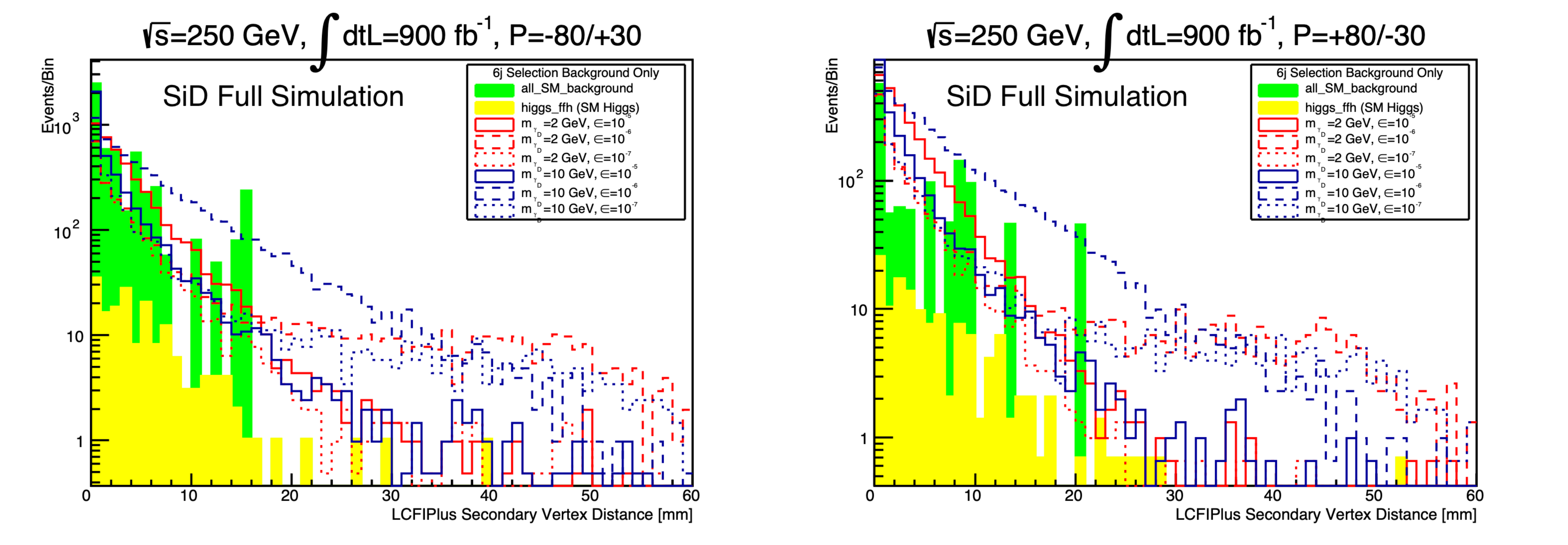}
\end{center}
\caption{Reconstructed secondary vertex distance for the  $e_{L}^{-} e_{R}^{+}$ (left) and $e_{R}^{-} e_{L}^{+}$ (right) samples at nominal beam polarization fractions. Secondary vertices associated to the $Z \rightarrow q\bar{q}$ candidate are discarded. The $6j$ signal selection has been applied to the backgrounds only. Signal is normalized with $BR(H \rightarrow \gamma_D \gamma_D)=0.01$.}
\label{fig:vertices}
\end{figure}

After full signal selection, we study the secondary vertices found by \texttt{LCFIPlus}. In each event, any secondary vertices within a $45^{\circ}$ cone around the $Z$ candidate jets are discarded. The remaining vertices are assumed to originate from the displaced $\gamma_D$ decays. See Figure \ref{fig:vertices} for the candidate dark photon secondary vertex distance from the interaction point after full $6j$ signal selection in the background samples. For the $4j2\ell$ and $6j$ selections, the effect of a provisional requirement for at least one secondary vertex reconstructed more than 2mm from the primary vertex ($r_{vtx}>2$~mm) can be seen in Table \ref{tab:bgyields}.

\begin{table}[p]
\begin{center}
\vspace{0.75in}
\begin{tabular}{|l|c|c|c|c|} \hline 
\textbf{Case $2j4\ell$ } & \multicolumn{2}{|c|}{$e^{-}_{L} e^{+}_{R}$ 900 fb$^{-1}$} & \multicolumn{2}{|c|}{$e^{-}_{R} e^{+}_{L}$ 900 fb$^{-1}$} \\
 Requirement & \texttt{higgs\_ffh} & \texttt{all\_SM}  & \texttt{higgs\_ffh} & \texttt{all\_SM}  \\ \hline \hline
All Events & $2.79\times 10^{5}$ & $2.6\times 10^{9}$ & $1.89\times 10^{5}$ & $2.53\times 10^{9}$ \\
$N_{e 10}+N_{\mu 10} \geq 4$ & $298$ & $1.37\times 10^{4}$ & $188$ & $1.06\times 10^{4}$ \\
$N_{trk} \in [10,40] \land N_{pfo} \in [20,80]$ & $116$ & $270$ & $81.6$ & $135$ \\
$N_{jet}=2$ & $116$ & $270$ & $81.6$ & $135$ \\
$-0.9 \leq \cos \theta_{jj} \leq -0.2$ & $50.5$ & $90$ & $44.7$ & $45$ \\
$75 \leq m_{jj} \leq 105$~GeV & $39.1$ & $0$ & $36.3$ & $45$ \\
$110 \leq m_{rec} \leq 140$~GeV & $16.5$ & $0$ & $9.77$ & $0$ \\
$-0.9 \leq \cos \theta_{\ell \ell} \leq -0.5$ & $15.5$ & $0$ & $8.37$ & $0$ \\
$110 \leq m_{4\ell} \leq 140$~GeV & $9.27$ & $0$ & $4.19$ & $0$ \\
$0.6 \leq m_{\ell^+ \ell^-} \leq 10.5$~GeV & $0$ & $0$ & $0$ & $0$ \\ \hline
\end{tabular}

\vspace{0.25in}

\begin{tabular}{|l|c|c|c|c|} \hline 
\textbf{Case $4j2\ell$}  & \multicolumn{2}{|c|}{$e^{-}_{L} e^{+}_{R}$ 900 fb$^{-1}$} & \multicolumn{2}{|c|}{$e^{-}_{R} e^{+}_{L}$ 900 fb$^{-1}$} \\
 Requirement & \texttt{higgs\_ffh} & \texttt{all\_SM} & \texttt{higgs\_ffh} & \texttt{all\_SM}  \\ \hline \hline
All Events & $2.79\times 10^{5}$ & $2.6\times 10^{9}$ & $1.89\times 10^{5}$ & $2.53\times 10^{9}$ \\
$N_{e 10}+N_{\mu 10}=2,3$ & $2.5\times 10^{4}$ & $1.93\times 10^{7}$ & $1.69\times 10^{4}$ & $1.82\times 10^{7}$ \\
$N_{trk} \in [10,40] \land N_{pfo} \in [20,80]$ & $2.02\times 10^{4}$ & $3.85\times 10^{5}$ & $1.36\times 10^{4}$ & $1.59\times 10^{5}$ \\
$N_{jet}=4$ & $1.93\times 10^{4}$ & $3.48\times 10^{5}$ & $1.3\times 10^{4}$ & $1.46\times 10^{5}$ \\
$-0.9 \leq \cos \theta_{jj} \leq -0.2$ & $1.01\times 10^{4}$ & $7.87\times 10^{4}$ & $6.83\times 10^{3}$ & $3.32\times 10^{4}$ \\
$75 \leq m_{jj} \leq 105$~GeV & $8.28\times 10^{3}$ & $3.58\times 10^{4}$ & $5.58\times 10^{3}$ & $1.64\times 10^{4}$ \\
$110 \leq m_{rec} \leq 140$~GeV & $3.34\times 10^{3}$ & $1.72\times 10^{4}$ & $2.25\times 10^{3}$ & $8.1\times 10^{3}$ \\
$-0.9 \leq \cos \theta_{\ell \ell,jj} \leq -0.5$ & $2.67\times 10^{3}$ & $1.14\times 10^{4}$ & $1.79\times 10^{3}$ & $5.44\times 10^{3}$ \\
$110 \leq m_{\ell \ell jj} \leq 140$~GeV & $692$ & $4.06\times 10^{3}$ & $422$ & $2.25\times 10^{3}$ \\
$0.6 \leq m_{\ell^+ \ell^-,jj} \leq 10.5,12.5$~GeV & $0$ & $0$ & $0$ & $45 \pm 45$ \\ \hline
$r_{vtx}> 2$~mm & $0$ & $0$ & $0$ & $0$ \\ \hline
\end{tabular}

\vspace{0.25in}

\begin{tabular}{|l|c|c|c|c|} \hline 
\textbf{Case $6j$} & \multicolumn{2}{|c|}{$e^{-}_{L} e^{+}_{R}$ 900 fb$^{-1}$} & \multicolumn{2}{|c|}{$e^{-}_{R} e^{+}_{L}$ 900 fb$^{-1}$} \\
 Requirement & \texttt{higgs\_ffh} & \texttt{all\_SM} & \texttt{higgs\_ffh} & \texttt{all\_SM} \\ \hline \hline
All Events & $2.79\times 10^{5}$ & $2.6\times 10^{9}$ & $1.89\times 10^{5}$ & $2.53\times 10^{9}$ \\
$N_{e 10}+N_{\mu 10} \leq 1$ & $2.44\times 10^{5}$ & $2.58\times 10^{9}$ & $1.65\times 10^{5}$ & $2.51\times 10^{9}$ \\
$N_{trk} \in [10,40] \land N_{pfo} \in [20,80]$ & $1.42\times 10^{5}$ & $1.11\times 10^{8}$ & $9.6\times 10^{4}$ & $6.41\times 10^{7}$ \\
$N_{jet}=6$ & $1.07\times 10^{5}$ & $6.89\times 10^{7}$ & $7.23\times 10^{4}$ & $4.26\times 10^{7}$ \\
$-0.9 \leq \cos \theta_{jj} \leq -0.2$ & $1.59\times 10^{4}$ & $8.15\times 10^{6}$ & $1.07\times 10^{4}$ & $2.13\times 10^{6}$ \\
$75 \leq m_{jj} \leq 105$~GeV & $1.28\times 10^{4}$ & $4.02\times 10^{6}$ & $8.64\times 10^{3}$ & $7.94\times 10^{5}$ \\
$110 \leq m_{rec} \leq 140$~GeV & $5.54\times 10^{3}$ & $2.08\times 10^{6}$ & $3.66\times 10^{3}$ & $3.32\times 10^{5}$ \\
$-0.9 \leq \cos \theta_{jj} \leq -0.5$ & $4.31\times 10^{3}$ & $1.55\times 10^{6}$ & $2.84\times 10^{3}$ & $2.46\times 10^{5}$ \\
$110 \leq m_{4j} \leq 140$~GeV & $2.45\times 10^{3}$ & $7.34\times 10^{5}$ & $1.59\times 10^{3}$ & $1.15\times 10^{5}$ \\
$0.6 \leq m_{jj} \leq 12.5$~GeV & $68$ & $6.86 \times 10^{3}$ & $51$ & $1.66 \times 10^{3}$ \\ \hline
$r_{vtx}> 2$~mm & $63$ & $1700 \pm 100$ & $47$ & $400 \pm 70$ \\ \hline
\end{tabular}
\end{center}

\caption{Background yields at each step in the signal $2j4\ell$ (top), $4j2\ell$ (middle) and $6j$ (bottom) selections. The \texttt{higgs\_ffh} sample is the SM Higgs background, not the signal dark photon sample. The \texttt{all\_SM\_background} sample includes all 2-, 3-, and 4-fermion SM backgrounds. Statistical uncertainties are suppressed until the last row, or if the uncertainty lies well below the implied precision.}
\label{tab:bgyields}
\end{table}

\subsection{Background Processes}

For the fully leptonic $2j4\ell$, $H \rightarrow \gamma_D \gamma_D \rightarrow 4\ell$ \emph{golden channel} case, all background is eliminated after candidate $\gamma_D$ mass selection. For both the $4j2\ell$ and $6j$ cases, most background arises from 4-fermion events $e^+ e^- \rightarrow WW,ZZ$ with semileptonic $WW,ZZ$ decays in the  $4j2\ell$ selection case and fully hadronic $WW,ZZ$ decays in the $6j$ selection case. More generally, these backgrounds are $e^+ e^- \rightarrow q\bar{q} \ell^+ \ell^-,q \bar{q}^{\prime} \ell \nu$ for the $4j2\ell$ case and $e^+ e^- \rightarrow q \bar{q} q^{\prime} \bar{q}^{\prime}$ for the $6j$ case, with on-shell $W,Z$ present or absent and at least one heavy quark $q=b,c$. Typically the 6-fermion signal is mimiced by 4-fermion backgrounds if gluon radiation from a quark occurs and creates an additional quark pair from gluon splitting. Such background events have a characteristic thrust signature, but this signature overlaps with signal events. 

In the $6j$ selection case alone, a secondary background arises from 2-fermion events $e^+ e^- \rightarrow b \bar{b} (\gamma), c \bar{c} (\gamma)$. Displaced vertices from heavy quark decays mimic the displaced $\gamma_D$ decay. Further, gluon radiation in these events with splitting to quark pairs can mimic higher fermion multiplicity final state events, and if the hadronic decay cones are wide these cones may be artificially split by the jetfinding. These hadronic 2-fermion events have a very high cross section due to radiative return to the $Z$. 

Higgstrahlung events with SM Higgs boson decays present a tertiary background in all signal selection cases, though it is eliminated in the case of the $2j4\ell$ selection and nearly eliminated in the case of the $4j2\ell$ selection. In the case of the $6j$ selection it presents a challenge. Because Higgstrahlung events with SM Higgs boson decays share the same characteristics as the signal events in the $Z$ and recoil mass selections, it is only distinguished by the Higgs boson decay characteristics. 

In the case of SM $H \rightarrow ZZ^{\star}$, the characteristics are nearly identical up to the distinguishing masses of the $Z,Z^{\star}$ and $\gamma_D$. The $Z^{\star}$ decay products must have very low invariant mass to mimic the dark photon $\gamma_D$ in the range of masses considered in this study, and both $Z$ bosons from the SM Higgs decay $H \rightarrow ZZ^{\star}$ rarely satisfy the candidate $\gamma_D$ mass requirements in the signal selections unless one (or both) is (are) misreconstructed. However, $Z^{\star} \rightarrow b \bar{b}, c \bar{c}$ produces displaced $B$ and $D$ meson decay vertices which can mimic the displaced $\gamma_D \rightarrow q \bar{q}$ vertex.

For both the $4j2\ell$ and $6j$ case, backgrounds from SM Higgstrahlung, 2-fermion and 4-fermion events can be somewhat reduced. The $b\bar{b}(\gamma)$ and $c\bar{c} (\gamma)$ background can be reduced with a veto on high energy photons. Tighter mass constraints on the hadronic $\gamma_D$ candidate in the signal selection can reduce it significantly, and a tighter constraint on the secondary vertex distance will further reduce it. Even for SM displaced vertices, however, a small tail at large distances from the interaction point may present an irreducible background.

\section{Summary and Outlook}

This work presents the first look at the sensitivity to detecting long-lived particles at the ILC. Using dark photons produced via the Higgs portal in Higgstrahlung events as an benchmark for light, weakly coupled particles with measurable lifetimes, we show that the ILC can expect to be competitive for the detection of dark photons with masses around 1~GeV. It is likely that this competitive edge extends below 1~GeV if it is possible to suppress background from photon conversions, and this is a promising area for future study. We also present the first full simulation of long-lived particles for SiD, and raise the interesting reconstruction challenges that may arise in signal events of this type in which there may be a reconstructable displaced vertex but no charged tracks at all from the primary vertex, in the case that the associated $Z$ boson decays to neutrinos.

\section*{Acknowledgements}

We thank David Curtin for his help with implementing the HAHM dark photon model, for providing us with the $\gamma_D$ width and BR calculations, and for enthusiastically answering our questions related to dark photon phenomenology. We also thank the Snowmass and Linear Collider communities who engaged with this study when we presented intermediate results. This material is based upon work supported by the U.S. Department of Energy, Office of Science, Office of High Energy Physics program under Award Numbers DE-SC0020244 and DE-SC0017996.

\bibliography{LOI-darkPhotons}% Produces the bibliography via BibTeX.

\end{document}